# Interaction-Region Decoupling through Structured Absorbing Potentials: A Framework for Scalable Time-Dependent Quantum Dynamics Calculations

Yuegu Fang,[1,3] Jiayu Huang,[1,2*] and Dong H. Zhang[1,3*]

[1] State Key Laboratory of Chemical Reaction Dynamics, Dalian Institute of Chemical Physics, Chinese Academy of Science, Dalian 116023, People's Republic of China

[2] Key Laboratory of Materials Modification by Laser, Ion and Electron Beams (Dalian University of Technology), Ministry of Education, Dalian 116024, China

[3] University of Chinese Academy of Sciences, Beijing 100049, People's Republic of China

`

*) To whom all correspondence should be addressed: jyhuang@dlut.edu.cn, zhangdh@dicp.ac.cn

**ABSTRACT**: Accurate quantum mechanical treatment of molecular reactions remains a longstanding challenge, especially for reactions involving deep potential wells and long-lived intermediate complexes. Here, we introduce an interaction region decoupling (IRD) strategy that incorporates structured absorbing potentials to dynamically partition the interaction region into reactant and product subspaces. The IRD framework integrates naturally with standard TDWP propagation schemes and enables the construction of region-specific basis sets, dramatically enhancing computational efficiency. Benchmark applications to the F + HD and O + OH reactions demonstrate that this approach achieves state-resolved accuracy while reducing computational cost by over two orders of magnitude. This strategy paves the way for routine quantum mechanical treatment of complex-forming four-atom reactions previously considered intractable.

## 1. INTRODUCTION

Chemical reactions involving deep potential wells play a key role in many areas of chemical physics, including combustion[1], atmospheric processes[2], plasma catalysis[3], and low-temperature chemistry[4]. In these systems, the formation of long-lived intermediate complexes leads to strong energy dependence and resonance features in the reaction probabilities. These effects cannot be explained by simple models and require accurate quantum mechanical calculations to fully understand the reaction dynamics. However, a fully quantum mechanical treatment of these reactions remains highly challenging due to the need to account for a large number of strongly coupled quantum states.

For deep-well reaction systems, time-independent close-coupling (TID-CC) methods become inefficient due to the rapid growth in the number of required basis functions and are typically limited to low total angular momentum[5-9]. An alternative is the time-dependent wave packet (TDWP) approach, which solves the time-dependent Schrödinger equation by propagating an initial wave packet to extract quantum dynamics information[10-13]. The TDWP approach avoids full matrix diagonalization and relies instead on efficient matrix-vector operations, making it more suitable for systems require large basis functions. To extract state-to-state reactive scattering information, TDWP methods have been developed using various coordinate representations. One class of approaches uses coordinate transformation: the wave packet is propagated in the reactant coordinates and transformed into product coordinates near the interaction region[14, 15]. While this offers accurate treatment of product states, the transformation is often numerically unstable.

Another class of approaches avoid coordinate transformation entirely by propagating the wavefunction in a single coordinate system. Among these, the most popular reactant-product decoupling (RPD) method[16-20] uses negative imaginary potentials to absorb the wave packet at a predefined "plane of no return", after which the absorbed flux is re-emitted in a separate propagation in product coordinates. More recently, the reactant-coordinate-based (RCB) method[21] extracts scattering amplitudes directly at a chosen surface in the product asymptotic region, eliminating the need for coordinate transformation altogether. Efforts have also been made to apply hyperspherical coordinates in TDWP dynamics[22]. These coordinates provide a unified treatment of the full configuration space and eliminate the need for coordinate transformation. However, the complexity of the kinetic energy operator in hyperspherical form significantly increases computational cost. As a result, its practical applicability has remained limited.

In all of the above methods, a single coordinate system is used throughout the interaction region. This is not very efficient, since the reactant and product regions differ greatly in geometry. For example, if the reaction Jacobi coordinates are used to represent the product region, the mismatch causes the wavefunction to become highly oscillatory or distorted near the product channels. This, in turn, requires very fine grids and large basis sets to resolve accurately. The problem becomes even worse when the wavefunction must be propagated deep into the asymptotic region, as is often the case in systems with deep potential wells or strong long-range interactions.

A major bottleneck in current TDWP approaches lies in the absence of an efficient coordinate representation for the interaction region. A promising strategy is to divide the interaction region into subdomains, each employing optimal coordinates. However, this requires wavefunction partitioning and real-time dynamic exchange of information between domains. In practice, strong quantum state coupling within the interaction region prevents such decoupling and often leads to numerical instabilities. To date, direct wavefunction exchange methods have proven possible only in asymptotic regions, where couplings are weak and the dynamics are comparatively simple.

In this work, we introduce an interaction region decoupling (IRD) approach based on structured absorbing potentials (SAPs). Inspired by the RPD method, SAPs are employed to absorb the wavefunction within the interaction region while maintaining continuity and avoiding spurious reflections. This enables the independent propagation of wavefunctions in locally optimal coordinate systems, avoiding both global transformations and coordinate mismatches. SAPs serve as smooth, non-reflective boundaries that preserve wavefunction continuity and allow for accurate and efficient flux transfer between subspaces. We demonstrate the effectiveness of this approach on two benchmark systems: the F + HD → D + HF reaction, which features sharp threshold resonances and strong long-range interactions, and the O + OH → $O_2$ + H reaction, which involves a deep potential well and multiple exit channels. In both cases, IRD achieves excellent agreement with benchmark results, achieving over two orders of magnitude improvement in computational efficiency compared to traditional approaches

## 2. METHODS

### 2.1 Wavefunction Decomposition and Propagation Scheme

The total time-dependent wavefunction $\Psi(t)$ is expressed as a sum of dynamically decoupled components:

$$\Psi(t) = \Psi_r(t) + \sum_p \Psi_p(t) \tag{1}$$

Here, $\Psi_r(t)$ describes the reactant component propagated in reactant Jacobi coordinates, and $\Psi_p(t)$ represents the wavefunction associated with the $p$-th product channel. Each component satisfies its own time-dependent Schrödinger equation and is propagated independently in its own Jacobi coordinate system. The reactant wavefunction evolves according to:

$$i\hbar \frac{\partial}{\partial t} \Psi_r(t) = \left(H_r - iV_r\right)\Psi_r(t) + i\sum_p V_p \Psi_p(t) \tag{2}$$

Here, $V_p$ is a smooth and negative imaginary potential that absorbs flux from the reactant before it enters the $p$-th product channel. The corresponding product wavefunction $\Psi_p(t)$ is propagated by solving an inhomogeneous time-dependent Schrödinger equation:

$$i\hbar \frac{\partial}{\partial t} \Psi_p(t) = \left(H_p - iV_p\right)\Psi_p(t) + iV_r \Psi_r(t) \tag{3}$$

The propagation of the wave functions is computed using the higher-order operator method[23] as

$$\Psi(t + \Delta t) = e^{-i\hat{H}\Delta t/\hbar} \Psi(t) \tag{4}$$

The total reaction probability can be calculated by evaluating the reactive flux at a dividing surface in the asymptotic region in the product region where potential energy is negligible.

This approach, as illustrated in Figure 1, partitions the full configuration space into non-overlapping regions, each treated in its optimal coordinate representation. The structured absorbing potentials (SAPs) act as non-reflective interfaces, ensuring smooth flux transfer while preserving wavefunction continuity. The absorbed wavefunction exchange between different coordinate is implemented by directly interpolate. In practice, the interpolation between $\Psi_r$ and $\Psi_p$ is performed at discrete time intervals rather than at every propagation step. Furthermore, the higher-order split-operator scheme is employed in the calculations to enable the use of larger time steps. In addition, due to the strongly exothermic nature of the deep-well reactions under study, the product flux rapidly separates from the interaction region, allowing the use of short absorbing regions. These strategies collectively reduce the computational cost associated with coordinate transformations and contribute only a negligible portion of the total CPU time.

**2.2 Coordinate Representation and Basis Functions**

In the following, we describe how to extend the IRD approach to atom–diatom reactive scattering systems using Jacobi coordinates for both the reactant and product channels. The Jacobi coordinate system is adopted to describe the relative motion between the atom and the diatomic molecule, with the Hamiltonian given by:

$$H = -\frac{\hbar^2}{2\mu_R}\frac{\partial^2}{\partial R^2} - \frac{\hbar^2}{2\mu_r}\frac{\partial^2}{\partial r^2} + \frac{(\hat{J}-\hat{j})^2}{2\mu_R R^2} + \frac{\hat{j}^2}{2\mu_r r^2} + V(R,r,\theta) \tag{5}$$

where $\mu_R$ and $\mu_r$ are reduced masses, $J$ is the total angular momentum, $j$ is the rotational angular momentum of the diatomic molecule, and $V(R, r, \theta)$ is the interaction potential. The full time-dependent wave function is expressed as:

$$\Psi^{JM\varepsilon}(R,r,\theta,t) = \sum_K D_{MK}^{J\varepsilon*}(\Omega) \sum_{n,v,j} F_{nvj}^K(t) u_n(R)\varphi(r) \tag{6}$$

where $M$ are the space-fixed projection of the total angular momentum, $\epsilon$ is the parity of the system, $K$ is the projection of $J$ onto the body-fixed (BF) axis, $D(\Omega)$ is the parity-adapted normalized Wigner rotation matrix, depending on the Euler angles $\Omega$, $\phi(r)$ is the vibrational eigenfunction of the initial diatomic reactant, $u_n(R)$ is the radial basis functions. The initial wave packet is constructed in the BF representation as

$$\Psi(t = 0) = G(R)\phi_{v_0,j_0}(r)| JMj_0K_0\varepsilon\rangle \tag{6}$$

where $G(R)$ is a is a Gaussian-shaped function describing a wave packet centered at $R_0$.

## 3. RESULTS AND DISCUSSION

As an initial application, we employ the IRD method to study the F + HD → D + HF reaction, a prototypical system characterized by strong long-range interactions. This reaction and its isotope systems has been extensively studied[24-32] and presents a stringent test for TDWP methods, particularly near the energy threshold where the HF($v'$= 3) product channel opens. In this regime, the reaction probability exhibits narrow resonance peaks and involves slowly propagating product wavefunctions, both of which pose computational challenges. All calculations are performed on the CSZ potential energy surface (PES)[33]. In this implementation, structured absorbing potentials $V_p$ are applied along the $R$ coordinate in the product Jacobi system, capturing the outgoing flux associated with each product channel. Similarly, a reactant-side absorber $V_r$ is introduced near the boundary to the product asymptotic region. The numerical parameters used in the calculations are carefully chosen to ensure full convergence. Detailed values are provided in Table 1.

Figure 2 presents a two-dimensional slice of the PES for the F + HD system, obtained by selecting the minimum energy configuration along the angular coordinate. As seen in Figure 2(a), the contour lines become highly distorted as the system approaches the product region, highlighting the limitations of using a reactant-based global coordinate system within the interaction region, as previously discussed. To address this, we introduce SAPs to decouple the wavefunction components across different regions. For simplicity, we employ an absorbing potential of the form, we use a simple absorbing potential defined as $V(x) = C_{abs}(x - x_0)/(x_{max} - x_0)$. Figures 2(b) and 2(c) show the SAPs constructed for the two product channels, with the corresponding distributions of $V_p$ and $V_r$ are shown as blue and red shaded regions, respectively.

indicated by blue and red shaded regions, respectively. For clarity, only those corresponding to the F + HD → D + HF reaction are shown in Figure 2(a). As illustrated in Figure 2, the decoupling region between the reactant and product coordinate spaces is located very close to the transition state. This placement allows the wavefunction to be propagated independently within the most suitable coordinate system for each region.

Figure 3 shows the calculated reaction probability for the F + HD($v = 0, j = 0$) → D + HF channel over a range of collision energies. When the collision energy exceeds 60 meV, the HF($v'=3$) product channel becomes energetically accessible. This threshold corresponds to the emergence of pronounced resonance structures, which are clearly resolved in the calculated probability curve. To evaluate the accuracy of our IRD method, we compare the computed reaction probabilities with those obtained from CC calculations using the ABC code[34]. The numerical parameters for the CC calculations were provided in Table 2. The results reveal that the IRD calculations closely reproduce the CC results, even in regions with extremely narrow resonance peaks. This strong agreement confirms the reliability of our approach.

Traditional TDWP calculations for this reaction are extremely demanding due to the need for dense spatial grids over extended regions. Previously, there are two decoupling methods have been proposed to solve this problem within the TDWP framework. One approach is based on hyperspherical coordinates[22], and the other employs a Jacobi-coordinate-based method combined with interaction-asymptotic region decomposition, referred to as JCB-IARD[35]. Although a direct comparison with the hyperspherical-coordinate-based method is difficult due to fundamental differences in coordinate structure and Hamiltonian form, the IRD method developed in this work achieves comparable accuracy while requiring only 21.3% of the grid points used in the hyperspherical method. In addition, it benefits from a more straightforward Hamiltonian, resulting in improved computational speed and easier integration into existing TDWP codes.

A more direct comparison can be made with the JCB-IARD method, as both approaches are implemented using the Jacobi coordinate system. According to previous reports, the JCB-IARD method required 255 grid points along the $R$ coordinate, 143 points in $r$, and 141 angular basis functions. In contrast, the IRD approach reduces these values to 50, 70, and 56, respectively. This represents a reduction to just 3.8 percent of the total basis set. Given that the computational cost of TDWP methods typically scales as $O(N^2)$, where $N$ is the total number of basis functions, this reduction yields an efficiency gain of approximately 700-fold. These results clearly demonstrate that the IRD method offers a practical and highly efficient solution for treating reactions involving long-range interactions within the TDWP framework.

To further assess the effectiveness and general applicability of the IRD method, we apply it to the O + OH → $O_2$ + H reaction. This system serves as a classic prototype of deep-well chemical reactions, featuring the formation of a long-lived $HO_2$ intermediate and exhibiting rich resonance structures in the reaction probability. Owing to its relevance in combustion chemistry and atmospheric processes, it has served as a critical benchmark for quantum dynamics methods[5, 36-41]. The complexity of the O + OH system arises from several factors: (1) the strong coupling between vibrational and rotational degrees of freedom in the intermediate region; (2) the presence of deep wells and multiple exit channels; and (3) the need to resolve dense transition-state structures. These features impose stringent demands on both the accuracy and efficiency of the IRD method.

Figure 4(a) presents a two-dimensional slice of the PES for the O + OH system, constructed in the same way as Figure 2(a). This reaction features two product channels: the $O_2$ + H formation channel and the O + OH exchange channel. For the $O_2$ + H channel, which is strongly exothermic, the outgoing flux is unlikely to return to the interaction region. As a result, we apply an absorbing potential $V_p$ in this channel but omit the returning potential $V_r$. In contrast, the exchange channel involves slow-moving wavefunctions components, making long-range interactions significant. To accurately decouple and propagate the corresponding wavefunction components, we apply SAPs on both sides of this region. The SAP distribution is shown in Fig. 4(a), and all numerical parameters are listed in Table 3.

Figure 4(b) presents the calculated reaction probabilities for the O + OH system. The results demonstrate excellent agreement with those obtained using the reactant-coordinate-based method, accurately capturing the narrow resonance features. Notably, the IRD method achieves this high level of accuracy while employing significantly fewer basis functions compared to single coordinate approaches. For direct comparison, the single-coordinate-based method uses 319 grid points along the $R$ coordinate, 127 points in $r$, and 110 angular basis functions. In contrast, the IRD method reduces these numbers to 99, 55, and 75, respectively, representing only about 9.2 percent of the total basis size required by the single coordinate approach. This reduction translates into a 118-fold increase in computational efficiency.

To assess state-resolved accuracy of the IRD method, we compute state-to-state reaction probabilities for the O + OH → $O_2$ + H reaction using the IRD method. The computed probabilities are compared to those obtained using the RPD method, which serves as a reference. Detailed values of the RPD calculations are provided in Table 4. Figure 4(a) shows an excellent agreement between the IRD and RPD results for individual product states ($v' = 0$, $j' = 1$) over a range of collision energies from 0.01 to 0.2 eV, indicating that the IRD method accurately describes the reaction dynamics at the state-to-state level. Figure 4(b) presents rotational state distributions of the $O_2$ product at vibrational quantum numbers $v' = 0$, 1, and 2. Again, the results obtained from the IRD method agree with those from the RPD method. These findings show that the IRD method is reliable in treating state-to-state complex reactive scattering systems.

The demonstrated accuracy and efficiency of the IRD approach in benchmark systems highlight its potential for broader applications. Notably, due to the prohibitive computational cost, fully quantum mechanical differential cross section calculations for complex-forming reactions involving four atoms, such as OH + CO and NaLi + NaLi, have not been reported previously. These systems present significant challenges due to deep potential wells, long-lived intermediates, and the need for extensive angular momentum coupling. The substantial computational savings enabled by the IRD approach now make such calculations feasible, marking a critical step toward the routine quantum treatment of complex-forming reactions in high-dimensional systems that were previously considered intractable.

## 4. CONCLUSIONS

We have developed the IRD method, a general and efficient framework for quantum reactive scattering involving deep wells and long-range interactions. By introducing SAPs within the interaction region, the IRD approach enables independent wavefunction propagation in locally optimal coordinate systems, avoiding the need for a global representation. This leads to significant reductions in basis size and computational cost. Applications to the F + HD and O + OH reactions show that the IRD method maintains high accuracy while achieving more than two orders of magnitude improvement in efficiency. While only the simplest form of SAPs is employed in this study, the concept can be readily generalized to more complex forms and extended to higher-dimensional systems. Notably, the enhanced scalability of IRD opens a path toward the routine quantum treatment of complex-forming four-atom reactions which have remained computationally inaccessible to date. Extending IRD to such systems represents a key direction for future research, with broad implications for advancing time-dependent quantum molecular dynamics in high-dimensional reactive systems.

Table 1. The numerical parameters of IRD calculations for the F + HD reaction

| | |
|---|---|
| Grid range (basis size) in F+HD channel | $R \in [0.0, 24.0], N_R = 255, N_{R,int} = 50$ |
| | $r \in [0.6, 8.0], N_r = 85,$ $N_{vib,int} = 50, N_{vib,asy} = 10$ |
| | $j_{max} = 55, Nj = 56$ |
| Grid range (basis size) in H+DF channel | $R \in [3.0, 21.0], N_R = 170, N_{R,int} = 31$ |
| | $r \in [1.0, 6.0], N_r = 60,$ $N_{vib,int} = 10, N_{vib,asy} = 5$ |
| | $j_{max} = 20, Nj = 21$ |
| Grid range (basis size) in D+HF channel | $R \in [3.5, 200.0], N_R = 2047, N_{R,int} = 31$ |
| | $r \in [1.0, 6.0], N_r = 60,$ $N_{vib,int} = 10, N_{vib,asy} = 5$ |
| | $j_{max} = 20, Nj = 21$ |
| Initial wave packet | $R_0 = 15.0, \delta = 0.8, E_0 = 0.05 eV$ |
| Total propagation time | 10 000 000 |
| Time step | $\Delta t = 30 \; with \; 6A8$ split operator |
| Steps of absorption for H+DF channel | 8 |
| Steps of absorption for D+HF channel | 5 |
| Matching plane for H+DF channel | $R_\infty = 15.0$ |
| Matching plane for D+HF channel | $R_\infty = 15.0$ |
| Absorbing potential in $R$ of F+HD channel | $n = 2, C_a = 0.001, R_a = 17.0$ |
| Absorbing potential in $r$ of F+HD channel (determined by H+DF channel) | $n = 2, C_a = 0.08,$ $R_a \in [4.0, 6.0] (in \; H + DF \; channel)$ |
| Absorbing potential in $r$ of F+HD channel (determined by D+HF channel) | $n = 2, C_a = 0.08,$ $R_a \in [4.5, 5.5] (in \; D + HF \; channel)$ |
| Left absorbing potential in $R$ of H+DF channel | $n = 2, C_a = 0.08, R_a \in [3.0, 4.0]$ |
| Absorbing potential in $R$ of H+DF channel | $n = 2, C_a = 0.02, R_a = 16.0$ |
| Left absorbing potential in $R$ of D+HF channel | $n = 2, C_a = 0.08, R_a \in [3.5, 4.5]$ |
| Absorbing potential in $R$ of D+HF channel | $n = 2, C_a = 2 \times 10^{-5}, R_a = 35.0,$ $C_b = 2 \times 10^{-4}, R_b = 155.0,$ $C_c = 2 \times 10^{-3}, R_c = 175.0,$ $C_d = 2 \times 10^{-2}, R_d = 195.0$ |

Table 2. Values of the input parameters in the CC calculations for the F + HD reaction

| Total angular momentum quantum number | $J = 0$ |
|---|---|
| Helicity truncation parameter | $K_{max} = 0$ |
| Maximum internal energy in any channel | $E_{max} = 3.2\ eV$ |
| Maximum rotational quantum number of any channel | $j_{max} = 32$ |
| Number of log derivative propagation sectors | $mtr = 6000$ |
| Maximum hyperradius | $R_{max} = 70.0\ bohr$ |

Table 3. The numerical parameters of IRD calculations for the O + OH reaction

| | |
|---|---|
| Grid range (basis size) in O+OH channel | $R \in [1.0, 32.0], N_R = 619, N_{R,int} = 99$ |
| | $r \in [0.5, 8.0], N_r = 110,$ $N_{vib,int} = 55, N_{vib,asy} = 10$ |
| | $j_{max} = 74, Nj = 75$ |
| Grid range (basis size) in H+O$_2$ channel | $R \in [5.0, 16.0], N_R = 219, N_{R,int} = 49$ |
| | $r \in [0.5, 4.5], N_r = 60,$ $N_{vib,int} = 11, N_{vib,asy} = 10$ |
| | $j_{max} = 35, Nj = 36$ |
| Grid range (basis size) in O+OH (exchange) channel | $R \in [3.5, 32.0], N_R = 559, N_{R,int} = 63$ |
| | $r \in [0.5, 4.5], N_r = 60,$ $N_{vib,int} = 7, N_{vib,asy} = 6$ |
| | $j_{max} = 30, Nj = 31$ |
| Initial wave packet | $R_0 = 17.5, \delta = 0.3, E_0 = 0.1 eV$ |
| Total propagation time | 700 000 |
| Time step | $\Delta t = 60 \ with \ 4A6a$ split operator |
| Steps of absorption for H+O$_2$ channel | 2 |
| Steps of absorption for O+OH (exchange) channel | 2 |
| Matching plane for H+O$_2$ channel | $R_\infty = 8.0$ |
| Absorbing potential in $R$ of O+OH channel | $n = 2, C_a = 0.001, R_a = 24.0$ |
| Absorbing potential in $r$ of O+OH channel (determined by H+O$_2$ channel) | $n = 2, C_a = 0.05,$ $R_a \in [5.0, 7.0] \ (in \ H + O_2 \ channel)$ |
| Absorbing potential in $r$ of O+OH channel (determined by O+OH (exchange) channel) | $n = 2, C_a = 0.08,$ $R_a \in [4.5, 6.0]$ $(in \ O + OH \ (exchange) \ channel)$ |
| Absorbing potential in $R$ of H+O$_2$ channel | $n = 2, C_a = 0.05, R_a = 11.0$ |
| Left absorbing potential in $R$ of O+OH (exchange) channel | $n = 2, C_a = 0.02, R_a \in [3.5, 4.5]$ |
| Absorbing potential in $R$ of O+OH (exchange) channel | $n = 2, C_a = 3 \times 10^{-3}, R_a = 12.0,$ $C_b = 3 \times 10^{-2}, R_b = 27.0$ |

Table 4. The numerical parameters of RPD calculations for the O + OH reaction

| | |
|---|---|
| Grid range (basis size) in O+OH channel | $R \in [1.0, 32.0], N_R = 619\ N_{R,int} = 319$ |
| | $r \in [0.5, 14.2], N_r = 127,$ $N_{vib,int} = 120, N_{vib,asy} = 10$ |
| | $j_{max} = 109, Nj = 110$ |
| Grid range (basis size) in H+$O_2$ channel | $R \in [9.0, 37.0], N_R = 559, N_{R,int} = 119$ |
| | $r \in [0.5, 4.5], N_r = 60,$ $N_{vib,int} = 11, N_{vib,asy} = 10$ |
| | $j_{max} = 30\ Nj = 31$ |
| Initial wave packet | $R_0 = 17.5, \delta = 0.3, E_0 = 0.1 eV$ |
| Total propagation time | 700 000 |
| Time step | $\Delta t = 60\ with\ 4A6a$ |
| Matching plane for H+$O_2$ channel | $R_\infty = 17.0$ |
| Absorbing potential in $R$ of O+OH channel | $n = 2, C_a = 0.001, R_a = 24.0$ |
| Absorbing potential in $r$ of O+OH channel | $n = 2, C_a = 0.05, r_a = 12.2$ |
| Absorbing potential in $R$ of H+$O_2$ channel | $n = 2, C_a = 0.01, R_a = 22.0$ |

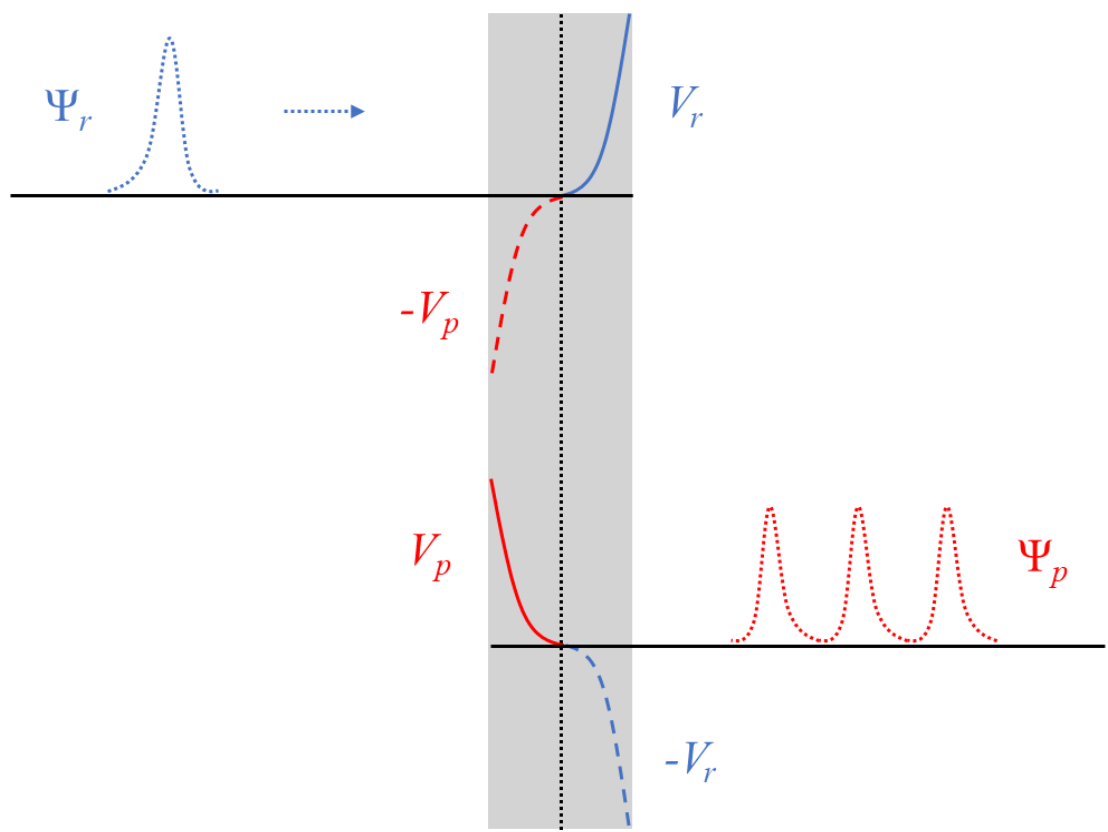

FIG. 1. Schematic representation of the interaction region decoupling strategy used in time-dependent wave packet calculations for an atom-diatom reactive scattering systems with deep potential wells. The total configuration space is divided into dynamically decoupled subregions corresponding to the reactant and each product channel. The reactant wave packet $\Psi_r(t)$ is propagated in the reactant Jacobi coordinates and partially absorbed by smooth negative imaginary potentials $-iV_r$ near the entrances of product channels. The absorbed flux is then used to initialize individual product wavefunctions $\Psi_p(t)$, each propagated independently in its own Jacobi coordinate frame. Similarly, $\Psi_p(t)$ is partially absorbed by an imaginary potential $-iV_p$, and the absorbed portion is re-emitted into $\Psi_r(t)$, enabling efficient round-trip propagation.

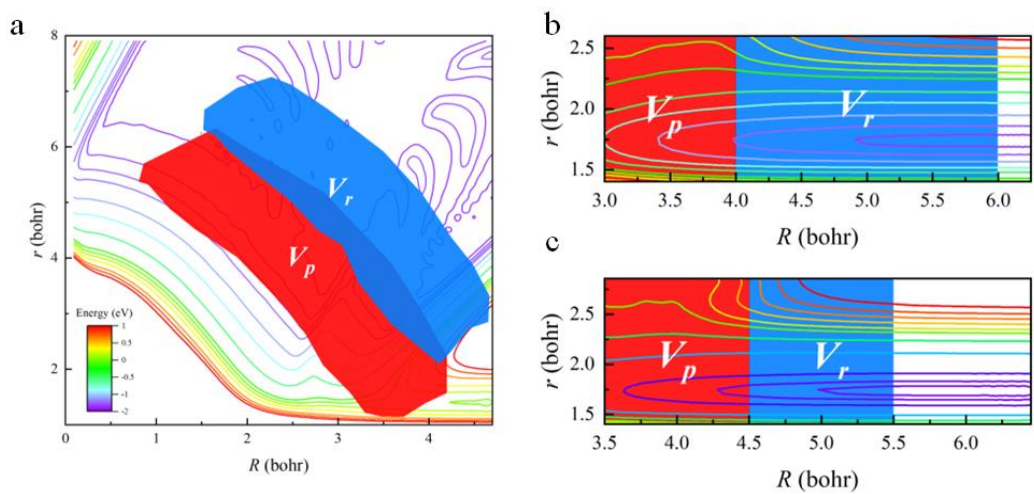

FIG. 2. Two-dimensional slice of the potential energy surface for the F + $H_2$ system, shown in (a) the reactant Jacobi coordinates, (b) the H + DF product Jacobi coordinates, and (c) the D + HF product Jacobi coordinates. Each plot corresponds to the minimum energy configuration along the angular coordinate. The structured absorbing potentials (SAPs) are indicated by the blue and red shaded regions. In the reactant Jacobi coordinates, only the SAPs associated with the F + HD → D + HF reaction channel is shown.

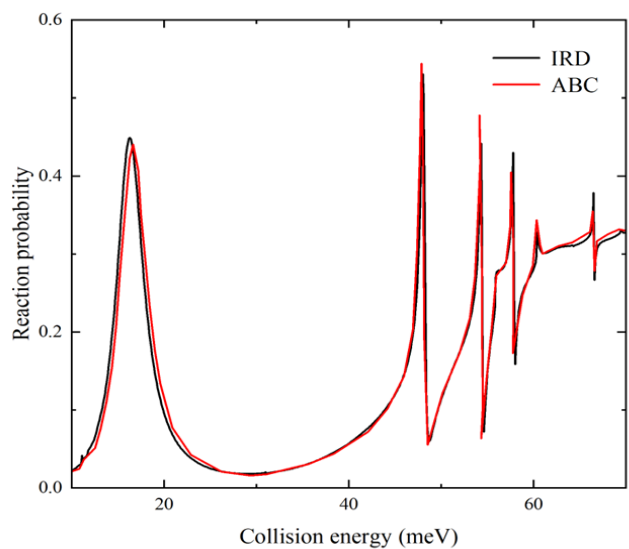

FIG. 3. Calculated reaction probability for the F + HD ($v = 0, j = 0$) → D + HF reaction as a function of collision energy. The interaction region decoupling (IRD) method accurately reproduces the benchmark results obtained from close-coupling calculations using the ABC code.

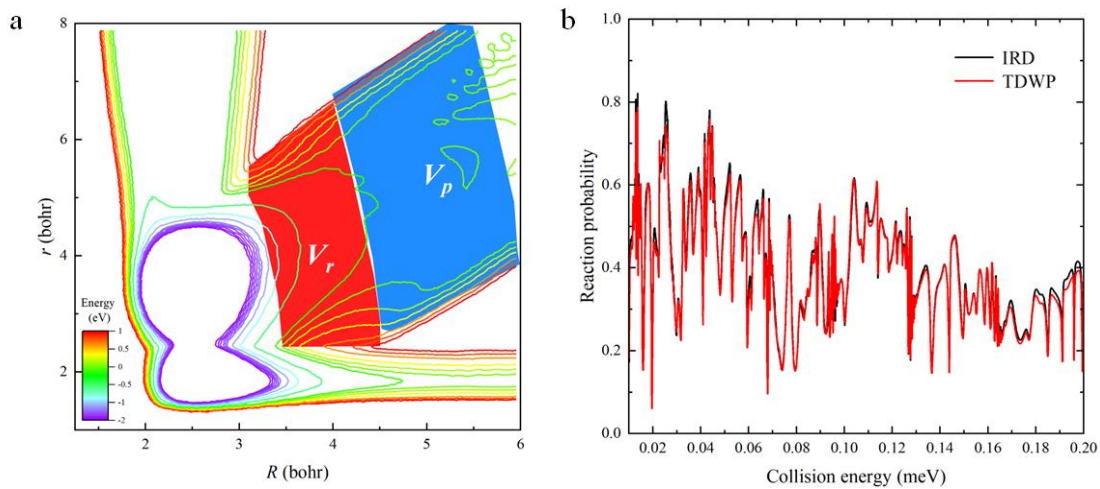

FIG. 4. (a) Two-dimensional slice of the PES for the O + OH system, shown in reactant Jacobi coordinates using the minimum energy path along the angular coordinate. Structure absorb potentials are introduced to decouple the long-range interaction in the exchange O + OH channel ($V_p$ and $V_r$). (b) Reaction probabilities for the O + OH($v = 0$, $j = 0$) → $O_2$ + H formation product channel calculated by the IRD method and the conventional time-dependent wave packet (TDWP) method.

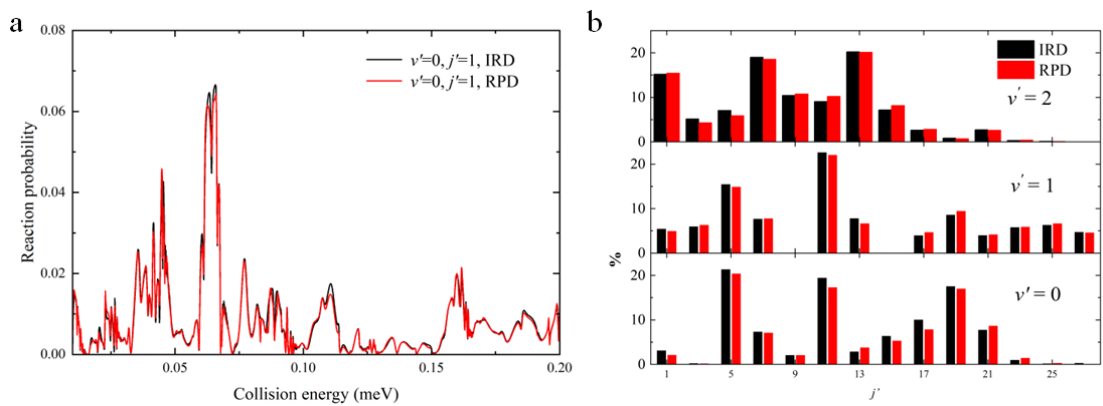

FIG. 5. (A) State-to-state reaction probabilities for the O + OH ($v = 0, j = 0$) → $O_2$ ($v' = 0, j' = 1$) + H reaction as a function of collision energy ranging from 0.01 to 0.2 eV, calculated using both the IRD and RPD methods. (B) Rotational state distributions of the $O_2$ product at vibrational quantum numbers $v' = 0$, 1, and 2, obtained from the IRD and RPD calculations.